30TH INTERNATIONAL COSMIC RAY CONFERENCE

# Gamma Air Watch (GAW): the electronics and trigger concept

P. Assis[1], G. Agnetta[2], P. Brogueira[1], O. Catalano[2], G. Cusumano[2], N. Gallì[2], S. Giarrusso[2], G. La Rosa[2], M. Pimenta[1], G. Pires[1], F. Russo[2], B. Sacco[2],

[1]*LIP, Av. Elias Garcia 14 -1, 1000-149 LISBOA, Portugal*
[2]*Ist. Astrofisica Spaziale e Fisica Cosmica, IASF-Pa / INAF, Via Ugo La Malfa 153, 90146 Palermo, Italy*
*Pedro.assis@lip.pt*

**Abstract:** GAW proposes a new approach for the detection and measurement of the Čerenkov light produced by GeV/TeV gamma rays traversing the Earth atmosphere which imposes specific requirements on the electronics design. The focal surface of the GAW telescope consists of a matrix of multi-anode photomultipliers. The large number of active channels (of the order of $10^5$) makes it basically a large UV sensitive digital camera with high resolution imaging capability. The limited amount of space available, due to the large number of channels, requires a compact design with minimal distance between the elements of the focal surface. The front-end electronics uses the single photoelectron counting technique to capture the Čerenkov light. The data acquisition is based on free-running data taking method. Self-triggering capability for each telescope is assured by detecting an excess of active pixels, in a 10ns time frame, inside overlapping trigger areas covering the whole focal surface. In this paper we describe the GAW electronics, as well as the trigger concept and implementation.

## The GAW concept

Traditional Imaging Atmospheric Čerenkov Telescopes (IACT) use large reflective optical systems associated with a PMT camera at the focal surface. These telescopes are designed to search for incoming γ-rays from a given source but have a small field of view (few degrees).

Gamma Air Watch [1,2] – GAW – is a "pathfinder" experiment to test the feasibility of a new generation of Imaging Atmospheric Čerenkov telescopes that join high flux sensitivity with large field of view (24º × 24º) capability. In traditional IACT design the size of the camera necessary to have a large FoV would produce a very large obscuration on the mirror. To overcome this problem GAW uses an innovative approach based on the use of a refractive optical system and a highly pixelated focal surface.

The refractive optical system is composed by a custom-made 2.13 m diameter Fresnel lens with a focal length of 2.56m. The lens is designed to have quite an uniform spatial resolution suitable to meet the Čerenkov imaging requirements up to 12º off-axis. The use of such a system allows overcoming the obscuration problem as well as the optical aberration for large input angles.

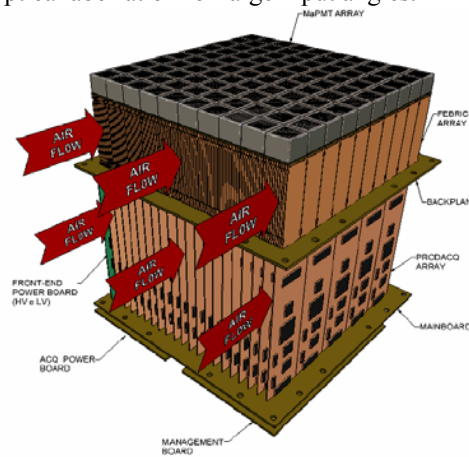

Figure 1: GAW Focal surface electronics

The focal surface detector of each telescope consists of a grid of 40 × 40 Multi-Anode Photomultipliers Tubes (MAPMT), with 64 anodes each, arranged in 8 × 8 matrix, operated in single photoelectron counting mode [3] instead of the charge integration method widely used in the IACT experiments. The total array of active channels, 102400 for each telescope, will record the Čerenkov image in a binary image with high granularity, which is fundamental in order to minimize the probability of photoelectrons pile-up within intervals shorter than the sampling time of 10 ns. In such working mode, the effects of electronics noise and the PMT gain differences are kept negligible, allowing to lower the photo-electron trigger threshold and, as a consequent result, to achieve a low telescope energy threshold in spite of the relatively small dimension of the Čerenkov light-collector.

## Electronics Layout

The GAW electronics has been designed to fully match the specific requirements imposed by the new proposed approach for the detection and measurement of the Čerenkov light produced by high-energy gamma rays traversing the Earth atmosphere. A large number of active channels constitute the focal surface of the GAW telescope making it basically a large UV sensitive digital camera with high resolution imaging capability. The GAW electronics design is based on single photoelectron counting method (front-end) and free running method (data taking and read-out)

### Focal Surface Detector configuration

GAW Focal Surface Detector (figure 2) is formed by an array of MAPMTs inserted in an electronic instrumentation UVIScope (Ultra Violet Imaging Scope) capable for conditioning, acquiring and processing a great number of high speed and high rate pulse signals.
The MAPMT used for GAW Focal Surface Detector is the Hamamatsu mod. R7600-03-M64 with 64 anodes arranged in an $8 \times 8$ matrix. The physical dimension of the tube section is $25.7 \times 25.7$ mm$^2$ (minimum true area is $18.1 \times 18.1$ mm$^2$), with length of $\sim$ 33 mm and weight of 30 g. The tube is equipped with a bialkali photocathode and a UV-transmitting window 0.8 mm thick. This ensures good quantum efficiency for wavelengths longer than 300 nm with a peak of 20% at 420 nm. The device has a Metal Channel Dynode structure with 12 stages, providing a gain of the order of $3 \times 10^5$ for a 0.8 kV applied voltage.

To quickly get a compact detection plane and assure as well a closed tubes assembling, the basic and repeatable parts of the UVIScope instrumentation (like the front-end and the acquisition signals) has been conceived in modular style. For that purpose the following two units was accomplished: a Front-End Brick unit (FEBrick), in order to accommodate just a single tube and a Programmable Data Acquisition unit (ProDAcq) for FEBrick unit signals acquisition.

### FEBrick (Front-End Brick)

FEBrick is a modular front-end unit, working in Single Photon counting mode, conceived for a single MAPMT and providing a low power active high voltage divider, 64 anodic channels and a high speed integration channel for the last dynode of the MAPMT. Each FEBrick module (Figure 2) acquires data from a line of 8 pixels. The FEBrick is composed by 8 bricks.

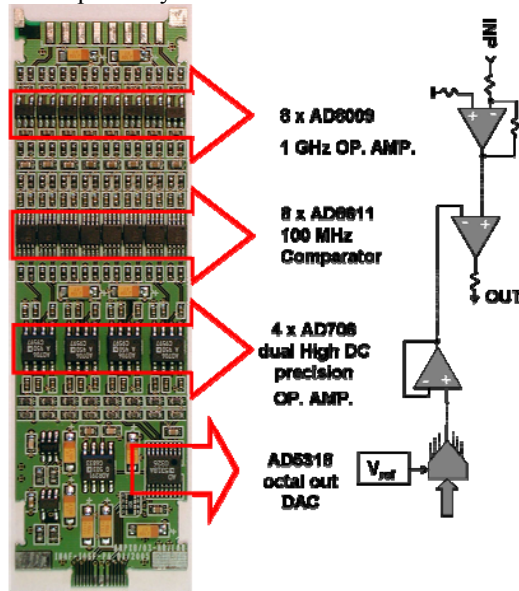

Figure 2: A module of the FEBrick.

The FEBrick (figure 3) develops on axis along the bottom of MAPMT as an appendix of identical section, which allows placing units side by side. The MAPMT is placed in a close contact with the input of the amplification channels.
FEBrick returns both digital photon location at the cathode surface and analogue information of the total charge detected from MAPMT. That image may be recorded by the acquisition system. FEBrick is equipped with:
• Tube insertion socket;
• Low power active high voltage divider;
• 64 anodic channels, operating in Single Photon Detection;
• 1 dynodic channel, operating as charge integrator;
• Digital reading temperature sensor.

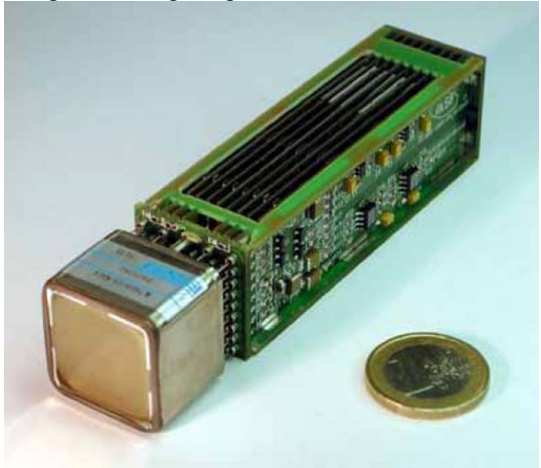

Figure 3: The FEBrick with a Multi-Anode PMT

All FEBrick units, required for the whole focal surface, are placed together on the top of a backplane while on the bottom are placed the ProDAcq units. Backplane is in charge to connect FEBrick unit signals to the relevant ProDAcq units.

## ProDAcq (Programmable Data Acquisition)

The ProDAcq (figure 4) unit is internally managed by a reprogrammable FPGA. Digital signals are recorded inside three memory banks for 192 Kword storage capacities. Input analogue signal may be sampled so fast and accurately according with the wiring combination of two ADC converters respectively running in AC and DC mode.
ProDAcq units are inserted on the bottom of backplane through which receive the signals of the relevant FEBrick units.
ProDAcq units are terminated on a mainboard equipped with Trigger and Timing Synchronization devices, instrumentation management, power supply and external host interface.

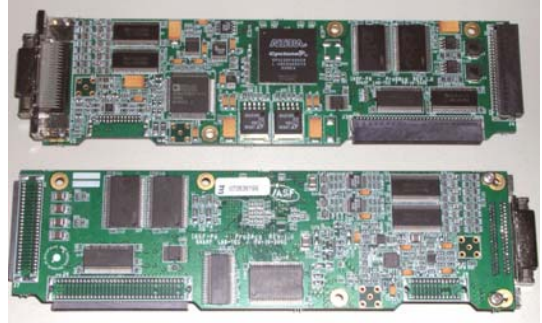

Figure 4: The ProDacq Board

## Backplane and Mainboard

The backplane should be designed both to contain all units required to compose the focal surface and to assure a suitable thermal dissipation of the whole assembling.
On the top of the backplane are assembled all FEBrick units while in the bottom are assembled ProDAcq units. Backplane connects the signals from FEBrick units to the relevant ProDAcq units.
On the lateral edges, on bottom side, the backplane houses the high voltage and low voltage section, the signal setting operating modality and control parameters section for all FEBrick units.
The mainboard is the base-plate that holds supplies and manages the focal surface electronic instrumentation. The mainboard contains ProDAcq unit assembling, Trigger and Timing Synchronization devices, instrumentation management, power supply and external host interface.
The motherboard contains a grid of FPGAs responsible for the second level trigger. Each FPGA receives data from $2 \times 2$ ProDacqs and is able to communicate with its FPGA neighbours.

## The Trigger System

The main requirement of the GAW trigger system is to be able to have a good efficiency at the lowest possible energy, keeping a high rejection power.
The noise counting rate of the telescope can be calculated knowing the average value of the background and applying the optical and geometrical characteristic quantities of the telescope. The average value of background rate <B> is 2200 photons m$^{-2}$ ns$^{-1}$ sr$^{-1}$. However, the option for a highly pixelated focal surface working at a high acquisition rate (100 MHz) allows having a small number of photoelectrons per pixel per time slot (~ 0.01). On the other hand this implies to be able to deal with a large number of acquisition channels at high speed.

The Čerenkov signal is intense and has a small time dispersion (<10 ns) and typically occupies an area of about 0.5º × 1º. Thus, the signal can be distinguished easily from the accidental background at reasonable energy threshold.

The concept of the GAW trigger system is based on three trigger levels. On the first trigger level the signals from the PMT pixels are subjected to a simple threshold and are transformed in digital signals. A pixel-on is defined here as a pixel which has a signal greater than the set threshold. On the second trigger level the focal surface is searched, online, for a given number of pixel-on inside all possible squares of 2 × 2 PMTs. On the third trigger level second level triggers are validated. It is also foreseen that this last trigger level will also decide the relevant data to be read from ProDAcq boards and communicate it to the data handling system.

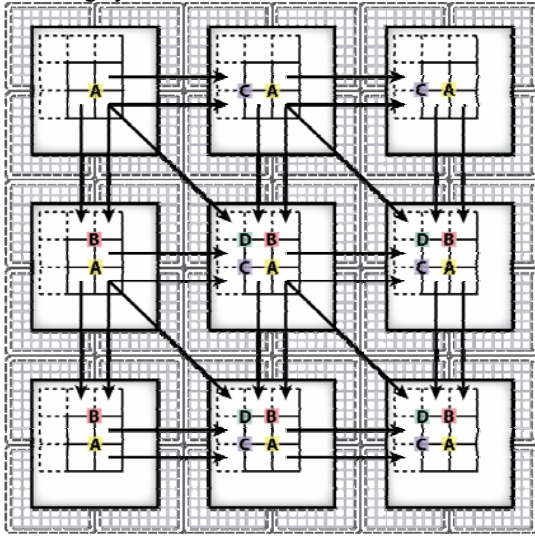

Figure 5: The second level trigger scheme.

This concept will be implemented in the electronics described before. The first trigger level is implemented on the front-end, namely as a fast discriminator. The second trigger level is implemented in the FPGAs present in the Mainboard. The third trigger level will be implemented on a FPGA that communicates with all the FPGAs on the Mainboard.

The second trigger level scheme is represented in Figure 5. The trigger operates online in a pipeline with three steps. On the first step each FPGA receives the number of pixel on in each ProDacq (and so in each PMT) attached to it. On the second step it transmits information about relevant PMTs to its neighbours on the right and on the bottom, receiving data from the top and left neighbours. This communication scheme allows, for each FPGA, to have the number of pixel-on for a set of 3 × 3 PMTS. On the third step each FPGA will search for a number of pixel-on greater than a programmable value in all possible squares of side 2 PMTs with the information available to that FPGA. If this condition is met a second level trigger is generated and passed to the third trigger level. With this scheme implemented on the whole focal surface the second level trigger system can search the whole focal surface for all squares 2 × 2 that meet the trigger condition.

## Conclusion

In this paper we have presented the electronic configuration and the triggering scheme for a new type of IACT. The electronic is based on the Single Photon Counting Technique. The triggering system relies on searching for a number of pixels with signal greater than a given threshold.

It is foreseen that the electronic and triggering system will be tested by the end of this year

## Acknowledgments

The authors from LIP acknowledge the support of FCT - Fundacão para a Ciência e Tecnologia, Portugal.